\documentstyle[twocolumn,seceq]{jpsj}
\newcommand{\ket}[1]{| #1 \rangle}
\newcommand{\bra}[1]{\langle #1 |}
\newcommand{\bracket}[2]{\langle #2 | #1 | #2 \rangle}

\newcommand{\Sqwpm}{{\cal S}^{+-}(Q,\omega)}

\def\runtitle{Dynamic Spin Correlation Function of the 
Alternating Spin Chain 
}
\def\runauthor{S. Mori, I. Harada and T. Tonegawa
}
\title{
Dynamic Spin Correlation Function of the Alternating Spin-1/2 
Antiferromagnetic Chain with Competing Interactions
}
\author{
Shigeyoshi {\sc Mori}, Isao {\sc Harada}$^{1}$ \footnote{E-mail: 
harada@cc.okayama-u.ac.jp} 
and Takashi {\sc Tonegawa}$^{2}$
}
\inst{
Department of Physics, Fukuyama University, Fukuyama 729-02,   \\
$^{1}$Department of Physics, Faculty of Science, Okayama University, 
Okayama 700 \\
$^{2}$Department of Physics, Faculty of Science, Kobe University, 
Rokkodai, Kobe 657
}
\recdate{\hspace{4cm}}
\abst{
Dynamic spin correlation function of the 
$S=1/2$ antiferromagnetic spin chain with alternating nearest neighbor 
and competing next nearest neighbor 
interactions is studied using the numerical diagonalization method.  
We find characteristics of the correlation function, which are the 
consequence of the dimer order in the ground state: 
(1) A typical singlet-triplet gap is observed and such a lowest-energy 
triplet state is found to form a bound state, 
showing strong peaks below 
a continuum band with weak but distinct upper and lower boundaries.  
(2) The intensity of the bound state shows a quite different 
wavevector-dependence from that of the corresponding intensity at the 
lower boundary in the 
usual antiferromagnet.  These features are examined by the variational 
calculation and are favorably discussed in connection 
with recent experimental results 
of neutron inelastic scattering in the inorganic spin-Peierls system, 
CuGeO$_3$.
}
\kword{
dynamic spin correlation function, dimer order, singlet-triplet energy gap, 
spin-Peierls system, CuGeO$_3$
}

\begin{document}
\sloppy
\maketitle
\section{Introduction}
It is realized recently that a spin gap due to the singlet spin dimer is 
rather common features in low dimensional quantum spin systems and plays 
important roles in the mechanism of the oxide superconductor.  
The spin-Peierls order, which is a consequence of the combined 
effects of the lattice distortion and the singlet dimer, is one of 
the highlights in this field and has attracted a renewed interest 
since the discovery of the inorganic Cu compound, CuGeO$_3$.~\cite{br}    

Because of large and high-quality samples, it becomes able to perform 
neutron inelastic scattering (NIS) for CuGeO$_3$,$^{2,3)}$
  which yields 
invaluable information on the dynamic properties of the spin-Peierls system. 
We believe that this is the most important progress in experiments comparing 
with the situation in organic materials about 15 years ago.~\cite{b}   
According to the 
experimental results, the dynamic spin correlation function seems to be, 
at least at a first glance, very similar to that of the typical 
one-dimensional antiferromagnet.  It is, however, realized several 
important differences: 
(1) At the wavevectors $Q=0$ and $\pi$ in units of $1/a$ 
where $a$ is the lattice constant in the uniform phase, 
it exhibits a typical gap.  
(2) In the whole wavevector space, bound state peaks seem to exist 
below the continuum band which has a rather distinct 
lower boundary.~\cite{al}  
(3) The intensity of the bound state decreases with decreasing $Q$ 
from $\pi$ but the decreasing rate is much slower than that in the 
typical one-dimensional antiferromagnet.~\cite{a}

It is the purpose of this paper to reveal characteristics of the 
dynamic spin correlation function of the one-dimensional quantum 
spin chain which has the dimer order in the 
ground state.~\cite{hmo}  We do not aim, at present, at making a 
quantitative comparison with the experimental results but want to 
clarify the characteristic features of the correlation function 
of the system described by the following Hamiltonian:
\begin{eqnarray}
{\cal H}&=&\sum_n [(1+(-1)^n\delta){\bf S}_n\cdot {\bf S}_{n+1}  
        + j {\bf S}_n\cdot {\bf S}_{n+2}],
\end{eqnarray}
where ${\bf S}_n$ represents the spin-$1/2$ operator at the $n$\/th 
site.  $\delta(1\geq \delta \geq 0)$ is the parameter describing the 
bond-alternation and $j(>0)$ is the strength of the next nearest neighbor 
(nnn) interaction divided by that of the nearest neighbor (nn) interaction 
which is set to be unity.   This Hamiltonian is 
believed to be appropriate for CuGeO$_3$,$^{6,7)}$
 although the 
parameter values for it are still controversial. 

The ground-state phase diagram of the system described by 
the Hamiltonian (1.1) is shown in Fig. 1,$^{8,9)}$
 where 
the space is spanned by the two parameters, $\delta$ and $j$, both of 
which lead cooperatively to the dimer ground state with an excitation 
gap.  As is seen in Fig. 1, the dimer order (D) takes place 
everywhere except for the $\delta=0$ and $j<j_{{\rm c}}
(\simeq 0.24)^{10,11)}$
 line, where the ground state is the usual 
gapless spin fluid (SF) state as in the case of $\delta=0$ and $j=0$.  
On the disorder line, $\delta=-2j+1$, which is shown by the dashed 
line in Fig. 1, the ground state is the direct product of the singlet 
dimers of nn spins so that a 
spin correlates only with the other spin in the dimer.  As the ground state 
on this disorder line is a prototype of the dimer order, 
we focus our attention in this paper mainly on the dynamic spin 
correlation function in the case of the dimer ground state on this 
disorder line, which starts from the 
independent dimer point, $(\delta=1, j=0)$, 
and ends at the point, $(\delta=0, j=0.5)$, called the Majumdar-Ghosh 
(MG) point.  It is worth noticing that the dimer ground 
states are doubly degenerate~\cite{ss} on the line, $\delta=0$ 
and $j>j_{\rm c}$, due to the translational symmetry of the Hamiltonian 
while a finite value of $\delta$ lifts the degeneracy.  This observation 
leads us to the fact that there is a domain wall excitation in the 
former case but not in the latter case, as will be discussed 
in the following sections.

In the next section, we describe our numerical method to reproduce 
the dynamic spin correlation function and present the result with 
some discussions.
The variational calculation is introduced in \S3 to extract a 
physical picture from the numerical result. 
The last section is devoted to our conclusion.

\section{Numerical Calculation}
In this section, we first summarize briefly the numerical method developed 
by Gagliano and Balseiro,~\cite{gb} which enables us to calculate 
the dynamic spin correlation function ${\cal S}^{zz}(Q,\omega)$, making 
this paper self-contained.  Then, the result of ${\cal S}^{zz}(Q,\omega)$ 
will be presented in a graphical form and its characteristics will 
be discussed.

We define ${\cal S}^{zz}(Q,\omega)$ by
\begin{eqnarray}
& &{\cal S}^{zz}(Q,\omega)  =  \frac{1}{2\pi}\int^{\infty}_{-\infty}{\rm d}t 
{\rm e}^{-{\rm i}\omega t}
                   \langle 0|  S^{z}_{-Q}S^{z}_{Q}(t) |0 \rangle \\
               & = & \frac{1}{\pi}\lim_{\epsilon\to +0} \nonumber \\
      &\times&
 {\rm Im} [\langle 0| S^{z}_{-Q}(\omega-i\epsilon +E_0 - {\cal H})^{-1}
                              S^{z}_{Q} |0 \rangle ],
\end{eqnarray}
where 
\begin{eqnarray}
S^{z}_Q & = &N^{-1/2}\sum_{n}\exp({\rm i}Qn)S^{z}_n,  \\
S_Q^{z}(t)&=&{\rm exp}({\rm i}{\cal H} t)S_Q^{z}{\rm exp}(-{\rm i}{\cal H} t),
\end{eqnarray}
Im[$\cdots$] means the imaginary part and $|0 \rangle$ denotes 
the ground state, which is obtained using 
the numerical diagonalization method with the standard Lanczos algorithm.  
Once the ground state
of a finite system is obtained, then ${\cal S}^{zz}(Q,\omega )$  can be 
calculated  
using the continued fraction expansion (CFE) for the resolvent in the
tridiagonal basis of the Hamiltonian with the initial vector
$S^{z}_Q|0\rangle$.  The tridiagonal expansion of the Hamiltonian can
be carried out numerically again with the Lanczos algorithm.  
Then, we obtain spectral weighs and poles of the
${\cal S}^{zz}(Q,\omega)$ with the CFE representation up to the order 
of 70.    
The convergence of the spectral weights and the pole positions are 
rather rapid in  the low energy excitations, while  it is worse in  
high energy excitations although they have only weak spectral 
weights.

In Fig. 2, we present our numerical results of the dynamic 
correlation function ${\cal S}^{zz}(Q,\omega)$ for $N=24$ for 
several parameter values on the disorder line together with 
the result of an exceptional case, the antiferromagnetic chain 
($j=0$ and $\delta=0$), for comparison.    

In Fig. 2 (a), we show a typical ${\cal S}^{zz}(Q,\omega)$ for the nn 
Heisenberg antiferromagnet.   The spectrum consists of the continuum band 
with the lower and the upper boundaries, which coincide with the known curves,
 $\omega_{{\rm L}}=(\pi/2) {\rm sin} (Q)$ and 
$\omega_{{\rm U}}=\pi {\rm sin} (Q/2$), 
respectively.  We note that the intensity is distributed within the band 
although the intensity at the lower boundary is high and seems to diverge 
especially at $Q=\pi$.  The result agrees with the one known 
for this system, showing the validity of our numerical calculation, 
although our spectra consist of the discrete poles due to 
the finiteness of the system size.

Now we proceed in cases where the ground state has the dimer order. 
We especially focus our attention on the perfect dimer cases, which occur 
on the disorder line as was mentioned in \S1.  The first example is the 
independent dimer point at which ${\cal S}^{zz}(Q,\omega)$ shows the delta 
function shape at $\omega=2$ without dispersion. It is easy to understand 
this since each isolated dimer has the singlet-triplet gap of 2 in our units. 
A bulk triplet state is obviously the state where one of $N/2$ 
nn singlet dimers in the ground state is excited to the triplet state.  
As $\delta$ decreases and $j$ increases, the degeneracy of these states 
is lifted and these states result in forming the triplet band, since 
the dimer propergates and spreads out due to the weak nn and nnn 
interactions.  
We show in Fig. 2 (b) an example of ${\cal S}^{zz}(Q,\omega)$ 
for this category, where the excited triplet states have a little 
dispersion.  It is interesting to note that the state which consists 
of a linear combination of 
such states with the wavevector $\pi/2$ is the exact eigenstate of 
the Hamiltonian as far as the parameters are on the disorder 
line.~\cite{cm}   

At another end of the disorder line, the MG point, 
it is known that the triplet 
states form a band whose elementary excitation is a domain wall connecting 
two degenerate ground states.~\cite{ss}  We note that the situation occurs 
only if $\delta=0$.  In Fig. 2 (d) the bound state predicted can be 
seen in the vicinity of $Q=\pi/2$ and seems to merge into 
the lower boundary of the continuum band at a certain wavevector.  
This observation is again consistent with the known result.~\cite{ss}

The most typical spectrum of the dynamical correlation function 
on the disorder line can be seen in Fig. 2 (c) and it also seems 
to be appropriate for the discussion of NIS in CuGeO$_3$.  
First of all, we want to note that the lowest pole for each 
wavevector is isolated from others.  Although the situation is 
smeared by the finite-size effect, we confirm this statement 
by the analysis of the size dependence of the positions and the spectral 
weights of such poles.
This suggests that the lowest poles form the bound state 
splitting off from the triplet continuum band, which has rather distinct 
lower and upper boundaries.  
%

Last point we would like to mention in this section is 
the wavevector dependence of the intensities at the lowest energy 
poles $\omega_{{\rm B}}$, which is shown in Fig. 3.   In 
this figure ${\cal S}^{zz}(Q,\omega_{{\rm B}})/{\cal S}^{zz}
(\pi,\omega_{{\rm B}})$ for $j=0.3$ is plotted as a function of $Q$.  
We note that the denominator ${\cal S}^{zz}(\pi,\omega_{{\rm B}})$ 
itself decreases rather rapidly when $\delta$ is introduced 
and the intensity is distributed to other wavevector modes.  
In addition, the higher energy tail shrinks as $\delta$ increases 
so that the intensity at the lowest energy mode seems to have 
a relatively high intensity.  Especially, at $Q=\pi/2$ the width 
becomes zero, so that the intensity can be comparable with that 
at $Q=\pi$.  This behavior of the intensities at the lowest 
energy poles is consistent with the experimental observation.

In the next section, we develop the variational calculation,
 emphasizing an important role of the local triplet dimer in the 
singlet dimer background, to make clear the nature of the triplet 
excitations studied in this section. 
%
%
\section{Variational Calculation}
In this section, we pursue a variational calculation for the excited 
triplet states, assuming the complete dimer singlet ground state, 
in order to gain insights into the nature of the 
triplet bound state as well as of the triplet continuum band, which 
have been shown in the previous section.  

Before going into the detailed calculation, we consider how to construct 
appropriate variational states.  The most plausible explanation for the 
low energy excited states in the system seems to be the one based 
on the domain wall scenario like in the case of the MG point.~\cite{ss}  
However, we note, when we introduce the alternation, 
the domain wall is no longer an elementary excitation of the system 
since the ground-state degeneracy is lifted by the alternation.  
We suppose the following two types of the states as our variational
 states:  (1) For the triplet bound state, we adopt the state in which 
one of $N/2$ singlet dimers in the dimer ground state is excited to the
 triplet dimer.  (2) For  the triplet band states, we consider the 
states in which two triplet dimers among $N/2-2$ singlet dimers 
couple into triplet. 
Note that the former state is nothing but an extreme state of the two 
domain wall states, where two domain walls are located on nn sites.  
It is clear that these states are the low-lying eigenstates in the
independent dimer limit ($\delta=1$,$j=0$). 

Based on the observation above, we build up the variational states. 
For convenience, we employ the two-spin states on a stronger 
bond of a nn pair as local bases of the Hamiltonian.  Then, 
the following symbols are 
introduced for the singlet state and the three components of the triplet 
state:
\begin{eqnarray*}
\ket{ \bullet_n } & \equiv & 2^{-1/2} 
      \ket{ \uparrow_{n}\downarrow_{n+1} - \downarrow_{n}\uparrow_{n+1}
}
\quad, \\
\ket{ 1_n } & \equiv & \ket{ \uparrow_{n}\uparrow_{n+1} } \quad,\\
\ket{ \circ_n } & \equiv & 2^{-1/2}
      \ket{ \uparrow_{n}\downarrow_{n+1} + \downarrow_{n}\uparrow_{n+1}
}
 \quad , \\
\ket{ {-1}_n } & \equiv & \ket{ \downarrow_{n}\downarrow_{n+1} }\quad,
\end{eqnarray*}
where $n$ denotes the site and is assumed to be the even number. 
For example, the dimer ground state is expressed 
by $ \ket{\rm D} = \ket{ \cdots\bullet\bullet\bullet\cdots } $.
We choose the following orthonormal states as 
the variarional bases for the low energy triplet states with the
wavevector $Q$:
\begin{eqnarray}
\ket{a_Q} & \equiv & (2/N)^{1/2}\sum_n {\rm e}^{{\rm i}Qm}\ket{a_m}
          \quad ,\\
\ket{b_{q_1 , q_2 } } & \equiv & 
C(q_{1}, q_{2})
\sum_{m,n} 
[{\rm e}^{{\rm i}(q_1 m +q_2 n)} - {\rm e}^{{\rm i}(q_1 n +q_2 m)}] 
\ket{b_{m,n}}
\quad ,
\end{eqnarray}
where $Q=q_1 + q_2$, and $C(q_{1}, q_{2})$ denotes the normalization
 constant, and
\begin{eqnarray}
\ket{a_m} &=& \ket{ \cdots\bullet\bullet 1_m \bullet\bullet\cdots }, \\
\ket{b_{m,n}} &=& \ket{ \cdots\bullet\bullet 1_m
\bullet\bullet\cdots\circ_{n}
\bullet\bullet\cdots }
\quad.
\end{eqnarray}
These $\ket{a_m}$ and $\ket{b_{m,n}}$ are depicted in Fig. 4.
It is notable that two triplet states, $\ket{b_{m,n}}$ and 
$\ket{b_{n,m}}$, are superimposed
antisymmetrically so as to produce the $S=1$ continuum states. 
In addition, we deal only with the triplet states with 
$S^z_{\rm t} \equiv \sum_n S^z_n = 1$, for the computational
convenience. 

Now, we deduce the effective Hamiltonian within the above restricted
Hilbert space.  To this end, we further take into account the effect 
of states consisting of three triplet dimers within the second order 
perturbation theory, as the number of the triplet dimers is not 
conserved in this system.   In other words, we integrate the process 
into the effective Hamiltonian so that the scattering matrix elements 
between the two triplet states, $\bra{b_{q_1, q_2}}{\cal
H}\ket{b_{q^{\prime}_1 , q^{\prime}_2}}$, and the diagonal elements, 
$\bracket{{\cal H}}{b_{q_1 , q_2}}$, are renormalized. Then, we
diagonalize numerically the effective Hamiltonian matrix and obtain 
the excitation spectrum and the dynamic
structure factor ${\cal S}^{+-}(Q,\omega)$ defined by
\begin{equation}
{\cal S}^{+-}(Q,\omega)  \equiv 
\frac{1}{2\pi}\int^{\infty}_{-\infty}{\rm d}t 
{\rm e}^{-{\rm i}\omega t}
                   \langle 0|  S^{-}_{-Q}S^{+}_{Q}(t) |0 \rangle, 
\end{equation}
where 
\begin{equation}
S^{\pm}_Q = S^{x}_Q \pm {\rm i}S^{y}_Q.
\end{equation}
Here we note that ${\cal S}^{+-}(Q,\omega)= 2{\cal S}^{zz}(Q,\omega)$ 
since the system has the rotational invariance.~\cite{x}

It is worth noticing that, in our variational calculation, 
the off-diagonal elements of the Hamiltonian 
$\bra{b_{q_1 ,q_2}}{\cal H}\ket{a_Q}$'s 
vanish at $Q=\pi/2$ so that $\ket{a_Q}$ becomes 
an exact eigenstate in this case. 
Reminding the fact that we assume the complete singlet  dimer 
ground state, 
which is true if the parameters are on the disorder line, 
we see that our variational calculation yields an exact 
solution at $Q=\pi/2$ and on the disorder line. 
It is also remarkable to note that eq. (3.5) suggests the spectral 
weights of $\Sqwpm$ depend on how much the state $\ket{a_Q}$ is 
mixed in
the excited states.   In fact, a contribution of the state $\ket{a_Q}$ 
seems to dominate $\Sqwpm$ as far as the ground state has the dimer
order. 

Next, we discuss the excitation spectra which are shown 
by the three dashed curves in Fig. 2 (b) and (c).  These are, 
hereafter, 
denoted as $\omega_{\rm B} (Q)$, $\omega_{\rm L} (Q)$, and 
$\omega_{\rm U} (Q)$ in the order of their energy, i.e., 
$\omega_{\rm B} (Q)<\omega_{\rm L} (Q)<\omega_{\rm U} (Q)$. The excitation
spectrum forms the continuum band in the region $\omega_{\rm L}
(Q)<\omega<\omega_{\rm U} (Q)$, while the lowest energy state is 
a bound state, 
whose dispersion curve $\omega_{{\rm B}}(Q)$ agrees well with the
lowest energy poles of ${\cal S}^{zz}(Q,\omega)$ for $\delta \ge 0.2$. 

It is quite reasonable that, the dispersion of all modes 
continuously tends to vanish with approaching the independent dimer 
point $\delta=1$ and $j=0$, where the ground state 
consists of the $N/2$ independent dimers.  In this limit, the 
first and the second excited states include, respectively, 
one local triplet 
and two local triplets in the singlet dimer background. 
Considering these, we conclude that the continuum band 
consists of two local triplets:  the upper and lower bounds are states 
where two triplets are located in the nn sites and in the distant 
sites, respectively.  As is seen in Fig. 2 the dispersion curve 
for the former coincides well with the numerical result while that of the 
latter does not.  We feel that, in order to obtain a reasonable result 
for the lower bound, the restriction of our variational calculation 
that the triplet is on nn sites must be released.  In this sense 
our result becomes worse when $\delta$ appraoches zero.

As was mentioned above, in the limit of the MG point, 
$(\delta=0,j=1/2)$, our perturbative approach fails to reproduce 
the continuum band obtained by the numerical diagonalization. 
For this case, the system possesses the full translational symmetry, 
giving rise to the degeneracy of the ground states.  In other words, 
an elementary excitation in this case is not such a local triplet state 
but a spin-1/2 
domain wall propagating on the lattice, as was shown by Shastry and 
Sutherland.~\cite{ss} In fact, we show in Fig. 2 (d) their result 
for the upper and the lower bound of the continuum band with the
 dash-dotted curve. In this limit, 
we see the bound state in the vicinity of $Q=\pi/2$, which is the 
same as the one discussed above, although it merges 
into the continuum band at the wavevectors $Q=0.36\pi$ and $0.64\pi$. 

In our variational approach, one additional bound state is observed in
the energy spectrum. it lies between the lowest bound state and the
lower boundary of the two triplet continuum as shown in Fig. 5. 
No correspondence is found in the result of the numerical diagonalization.
 But it has only the quite small contribution to the scattering intensity. 
Perhaps it is caused by the strong restriction for the Hilbert space in our
variational calculation, i.e. , each pair of spins that forms the local
triplet state are bound more gradually in space, not bound strongly on
the nearest neighbour sites.

In conclusion, we see that our variational calculation gives us a
reasonable 
explanation for the dispersion for the bound state as well as 
the continuum band obtained in the numerical calculation except 
for its lower boundary.  
Now, we turn to the wavevector dependence of their 
intensities on the lowest energy states. The behavior of the intensities 
coincides with that of the numerical result, as shown in Fig. 6, not only
qualitatively 
but also quantitatively, although this coincidence becomes worse when
the parameters tend to the MG point.
Decreasing $\delta$ along the disorder line, we find the spectral 
weight of the lower energy mode decreases while that of the higher 
energy mode gets the intensity, except at $Q=\pi/2$ where only 
the lower energy mode has the intensity.  This 
tendency is again consistent with the result of the numerical
calculation. 
%
%
%
\section{Conclusion}

In conclusion, we have found that in the $S=1/2$ antiferromagnetic chain 
with alternating nn interactions and competing nnn 
interactions, the dynamic correlation function shows characteristic 
features due to the dimer order in the ground state:  (1) The ground state 
is the disordered state so that the static structure factor, which is 
obtained by an integration of  ${\cal S}^{zz}(Q,\omega)$ with respect to 
$\omega$, has a 
relatively high intensity even at wavevectors other than $Q=\pi$.  
This is marked contrast to the case in the spin-fluid state.  (2) 
The width of ${\cal S}^{zz}(Q,\omega)$ at $Q=\pi/2$ becomes 
narrow as the disorder line is approached and eventually reaches zero 
at the disorder line.  (3) The lowest pole for each wavevector splits 
off from the continuum band.  We elucidated the nature of these states 
by the variational calculation assuming the triplet states based on 
the physical picture, and confirmed the facts mentioned above.   

All the behavior observed in our results is consistent with the 
experimental observation of NIS.$^{2,3)}$
   This suggests that 
the parameters of CuGeO$_3$ are close to those on the disorder line.  
We believe that the reason why the NIS result shows very different 
characteristic behavior from what we expect for a small dimerization 
in the nn spin-Peierls system is a manifestation of the important role 
played by the rather large competing nnn interaction.  
We have not aimed at comparing quantitatively our theoretical result 
with the experimental result of CuGeO$_3$ in this paper, because we feel 
that the following issues should be considered for a quantitative 
comparison. 

First we should throw light on the effect of the interchain interactions, 
since they are not so weak and said to be only one order of magnitude 
smaller than the intrachain interactions.  We believe, however, that 
the effect is reduced by singlet fluctuations caused by the rather large 
nnn interactions.  
Second a possibility that the dimerized lattice distortion can relax 
according to the spin state should be considered.  If the lattice distortion 
follows completely the spin behavior, our Hamiltonian is inadequate 
in the sense that the distortion is treated as rigid.  In such a case 
a soliton state like the one discussed for the magnetic ground states in 
high fields$^{16,17)}$
 may be a proper excited state.  In this situation, 
information for the phonon is highly desired.  In any case this is 
an interesting as well as a challenging problem for our future study. 
Lastly, the condition for the bound state to exist should be clarified 
from both experimental and theoretical sides.

At the final stage of our study, we have been notified that Yokoyama 
and Saiga~\cite{ys} have studied the same problem using the same method.  
They claimed for CuGeO$_3$ rather large value of $j$ keeping a small 
value of $\delta$, which is consistent with our conclusion in the sense 
that, in order to see characteristic features of the spin correlation 
function, the system should be close to the disorder line.    
We hope that these calculations stimulate further experimental work 
of NIS to elucidate the nature of the bound state as well as the continuum 
band, which enable us a quantitative comparison with theoretical calculations.
\section*{Acknowledgments}
The authors would like to thank Professor M. Arai and Mr. M. Fujita of KEK 
for valuable discussions concerning their experiment, which have 
encouraged them throughout this work.  Thanks are also due to 
Professor H. J. Mikeska of Hannover University for his useful 
comments.
Our computer programs for the numerical diagonalization of the Hamiltonian 
is based on the subroutine package, TITPACK Ver.2, coded by Professor H.
 Nishimori.
The computation in this work has been done using the facilities of 
the Supercomputer Center, Institute for Solid State Physics, University 
of Tokyo. 
\newpage

\section*{\large\bf Figure Captions}

\begin{itemize}
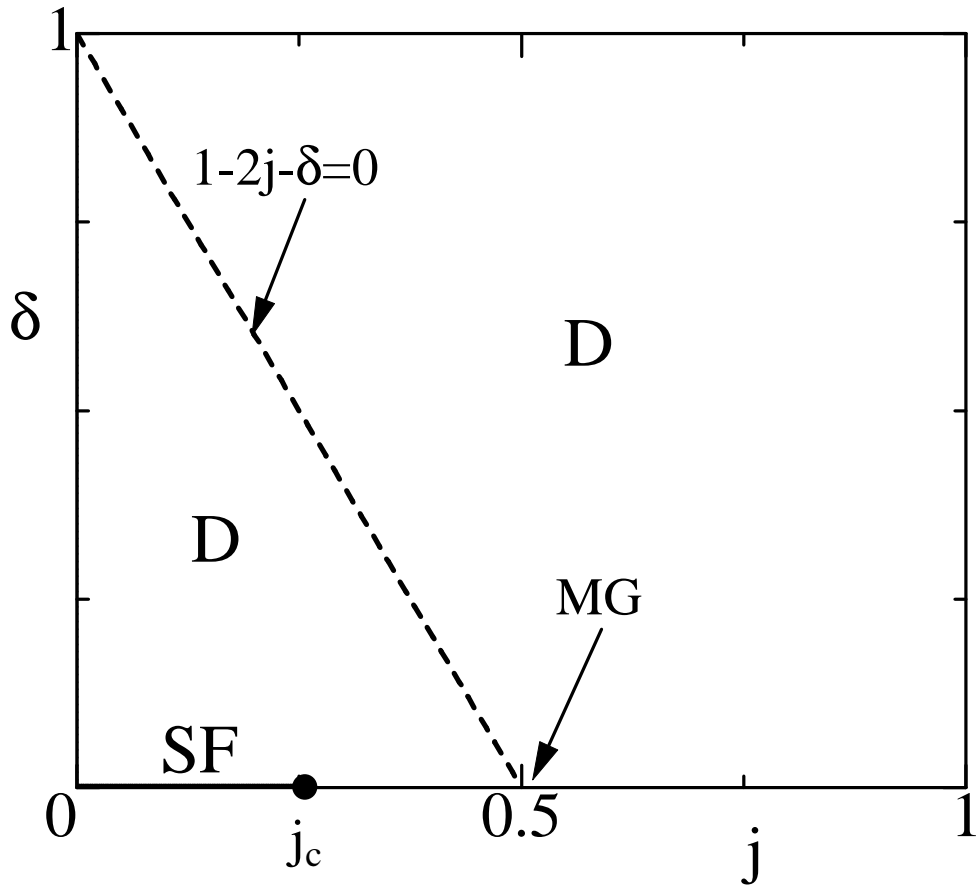
\item[Fig. 1]
Ground-state phase diagram in the parameter space spanned by the 
alternation parameter $\delta$ and the competing next nearest neighbor 
interaction $j$.  The solid line shows the spin-fluid (SF) phase without 
a spin gap.  Other region is the dimer (D) phase with the spin gap. 
The disorder line is denoted by the dashed line and the Majumdar-Ghosh 
(MG) point is indicated.
\vspace*{1cm}
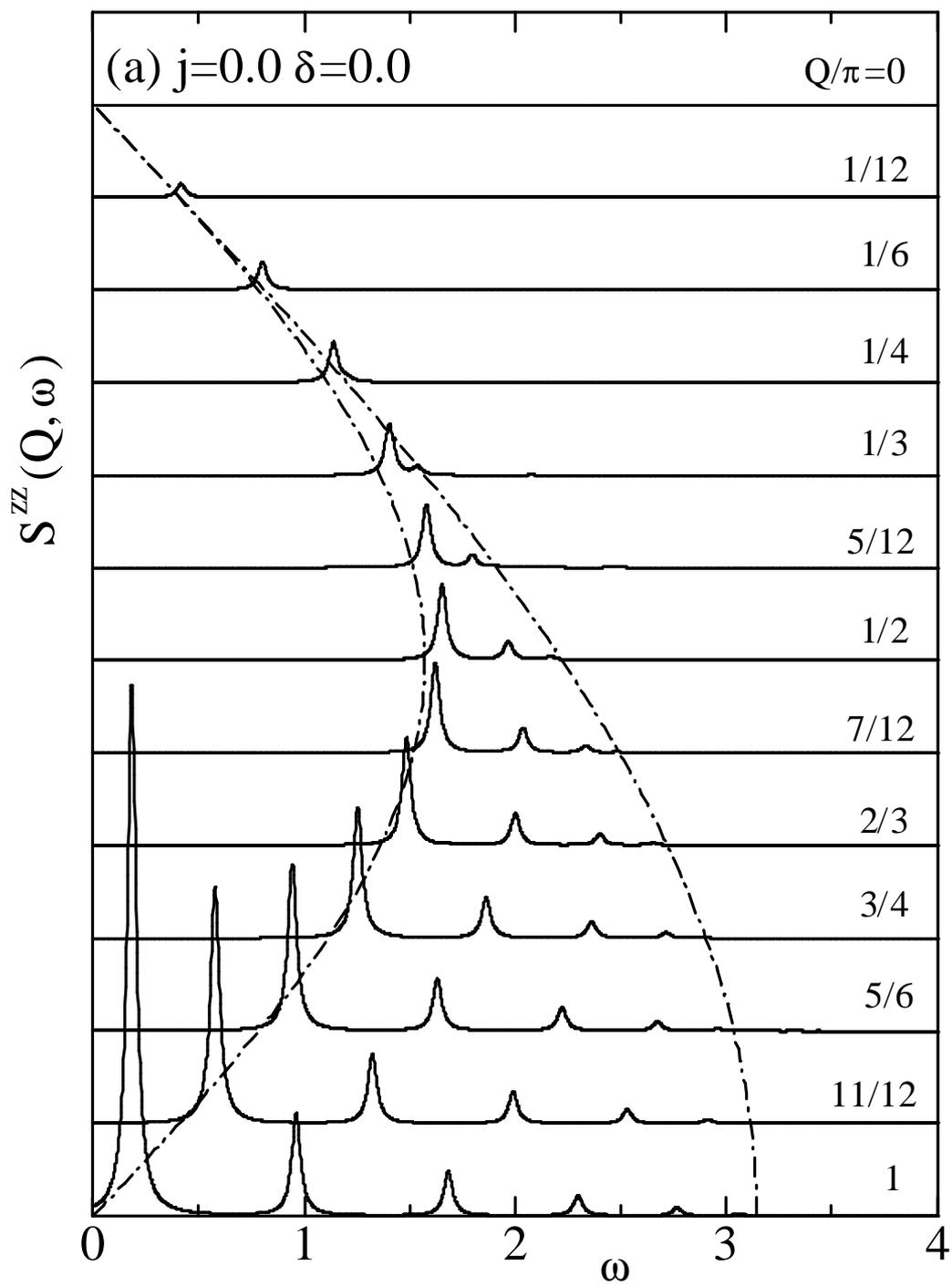
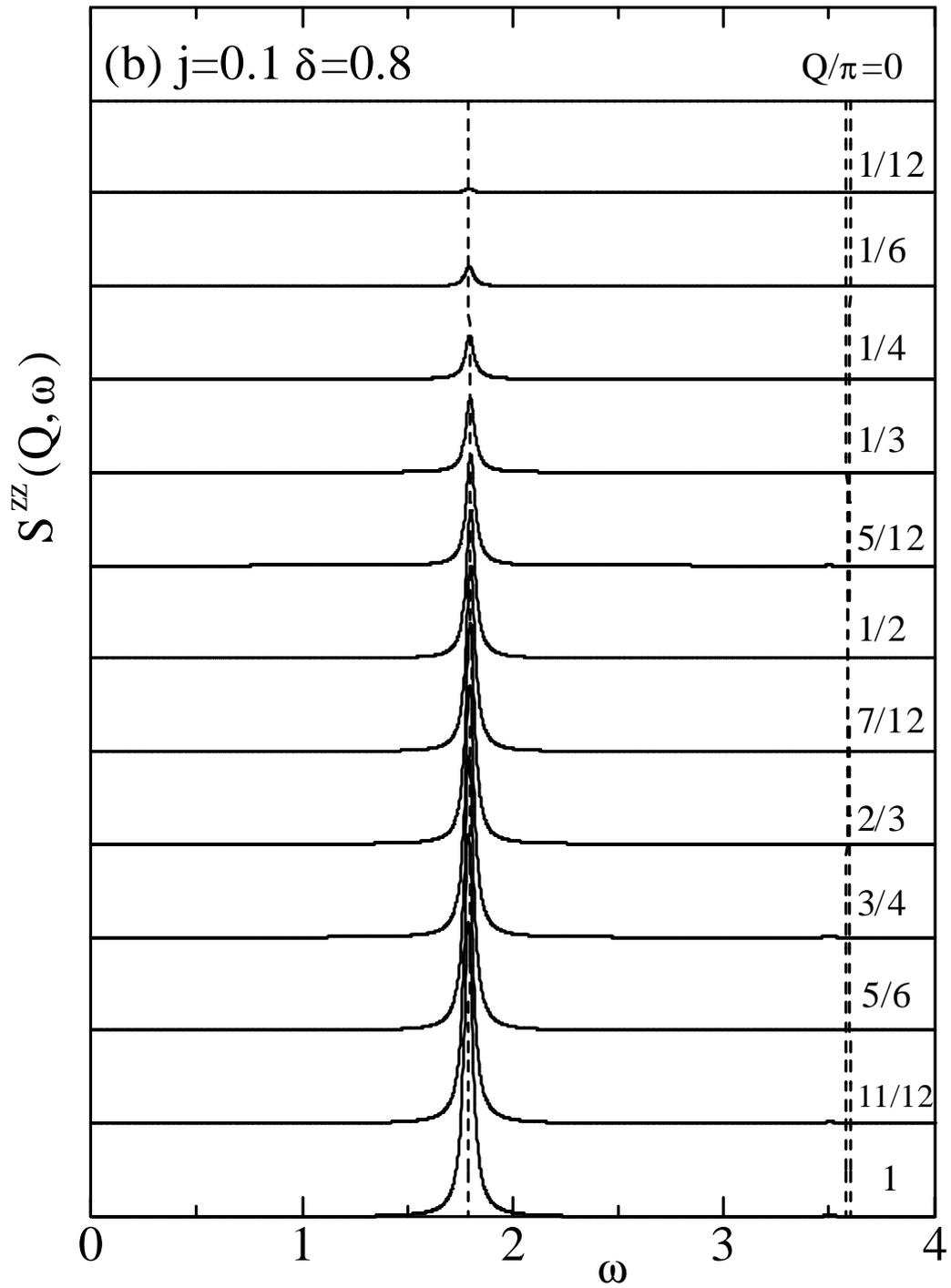
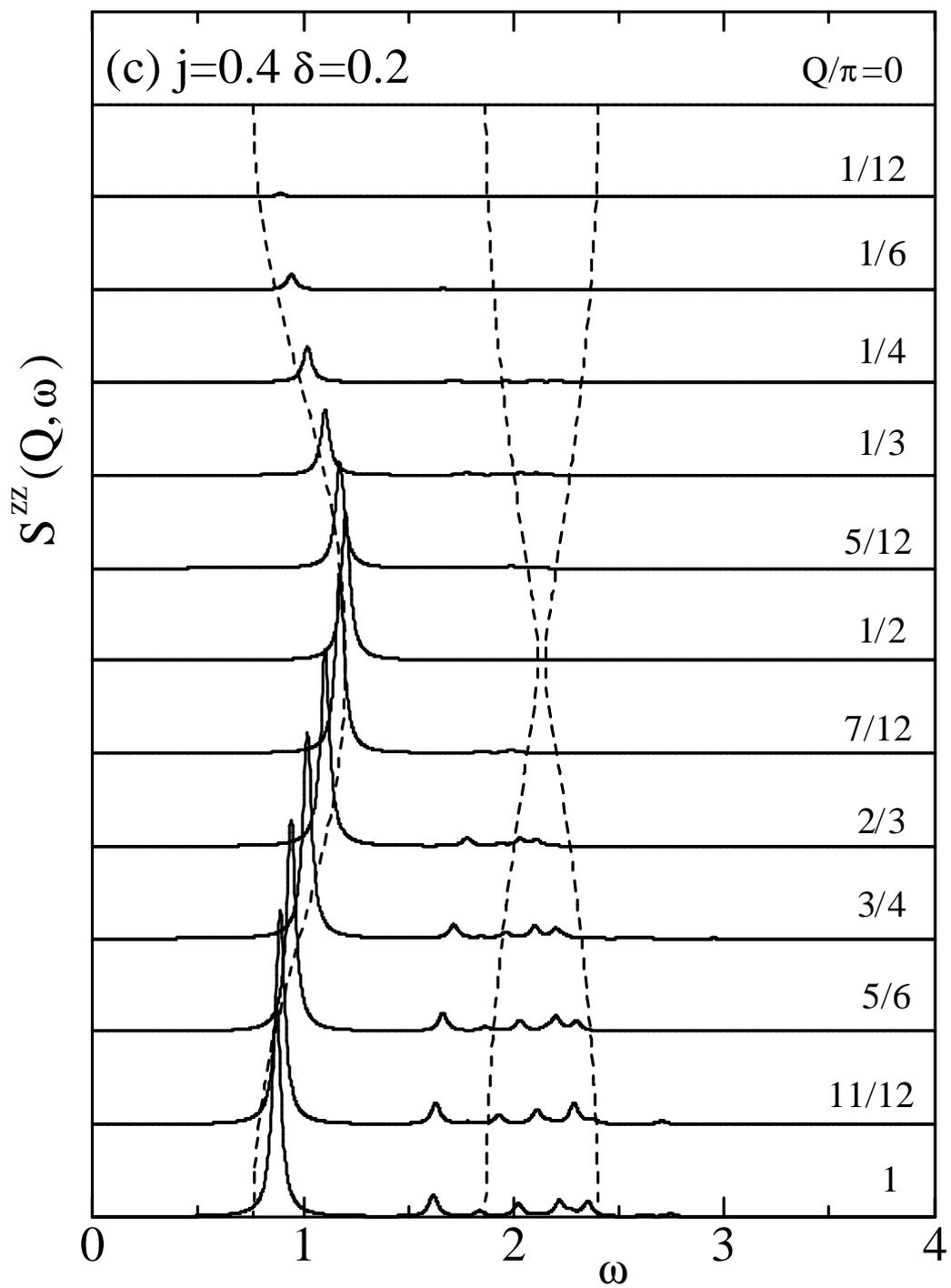
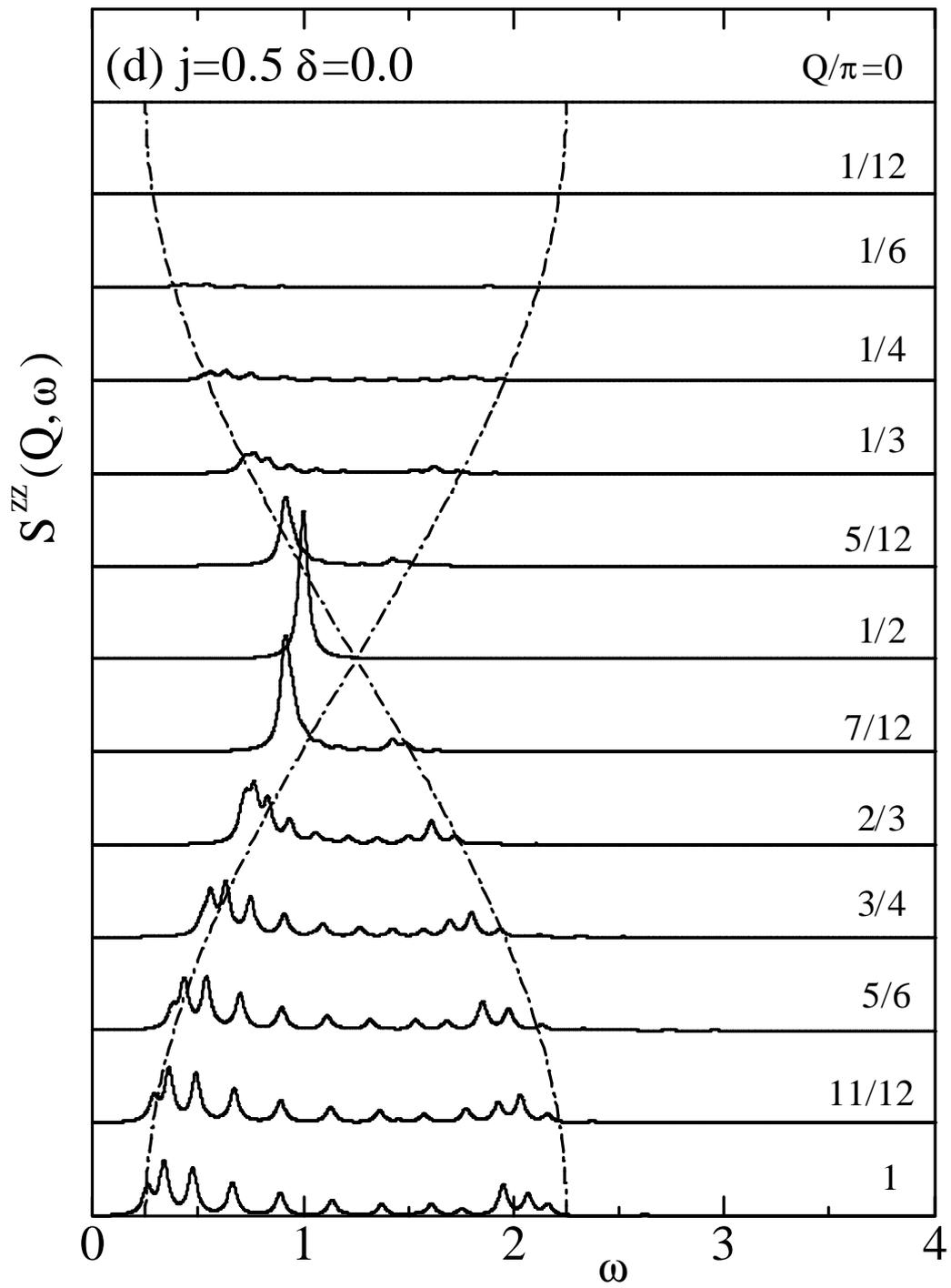
\item[Fig. 2]
Dynamic spin correlation function of the finite spin system ($N=24$), 
${\cal S}^{zz}(Q,\omega)$, for representative values of parameters, 
$\delta$ and $j$. In (a) the dash-dotted curves show the rigorous upper and 
lower boundaries of the triplet band, while in (b), (c) the approximate
 dispersion curves of the traveling triplet dimer in the singlet-dimer
 background are represented by the dashed curves. (See text.) In (d), the
 dash-dotted curves show the two-particle scattering continuum of the 
$S=/2$ domain walls~\cite{ss}
\vspace*{1cm}
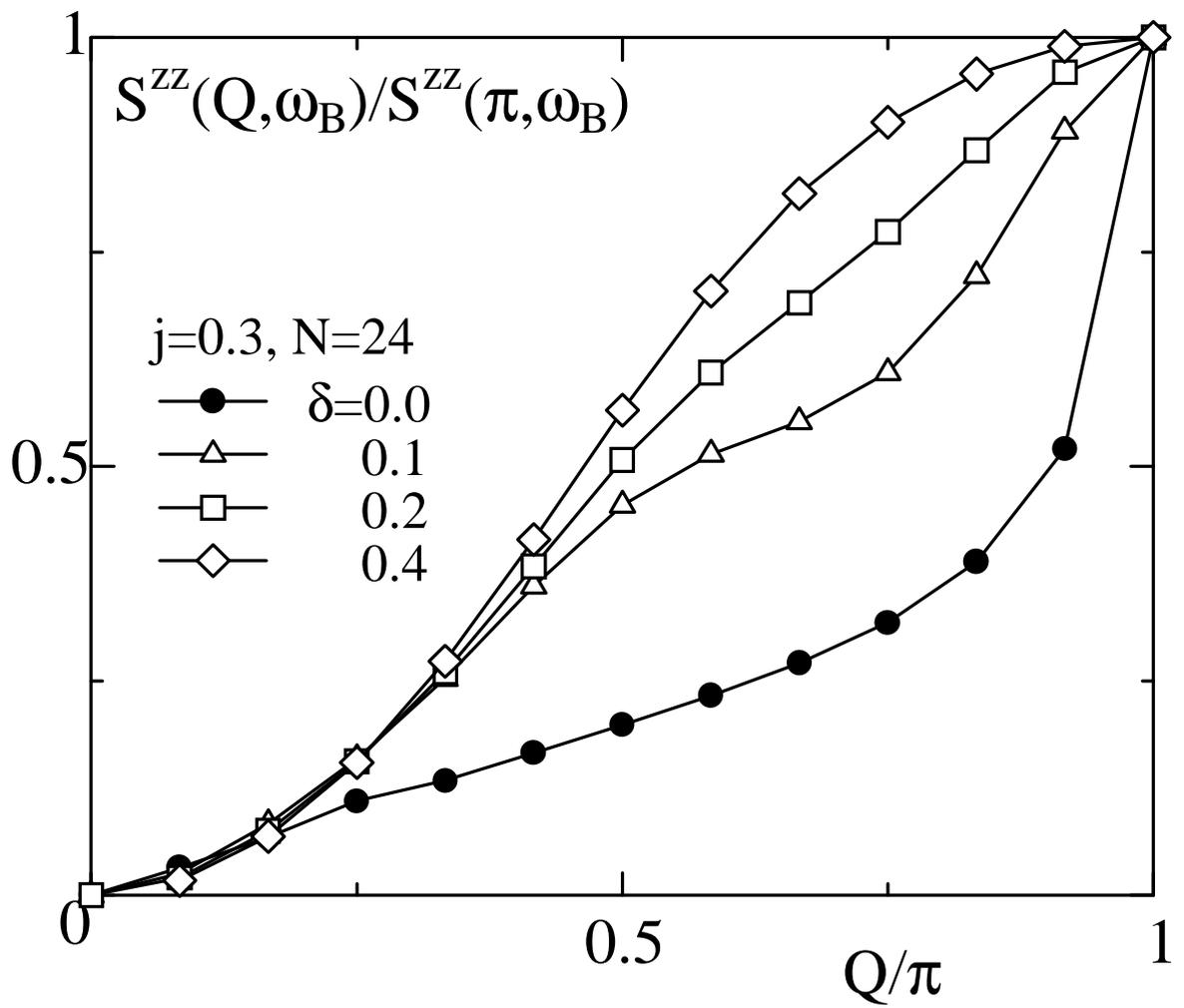
\item[Fig. 3]
Intensity ratio of the lowest energy poles, \\
${\cal S}^{zz}(Q,\omega_{{\rm B}})/{\cal S}^{zz}(\pi,\omega_{{\rm B}})$,
 for $j=0.3$ 
and representative values of $\delta$ as a function of the wavevector $Q$.
\vspace*{1cm}
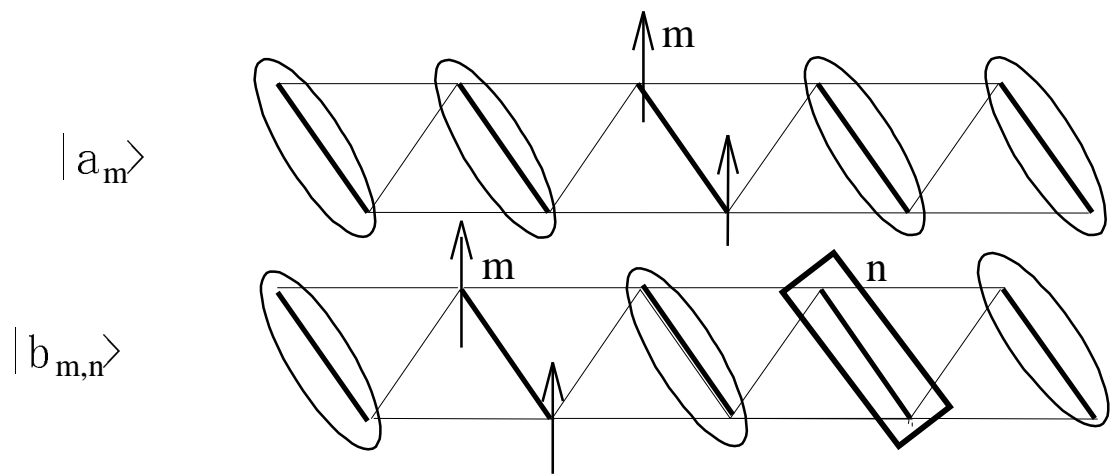
\item[Fig. 4]
Two types of triplet states adopted to construct the trial wave functions. 
In each state, the two arrows form the triplet state with 
$S^z_n + S^z_{n+1} =1$ while the solid rectangle denotes the triplet pair
 with $S^z_n + S^z_{n+1} =0$.  The solid ellipse shows the singlet pair 
which is the same as in the ground state.
\vspace*{1cm}
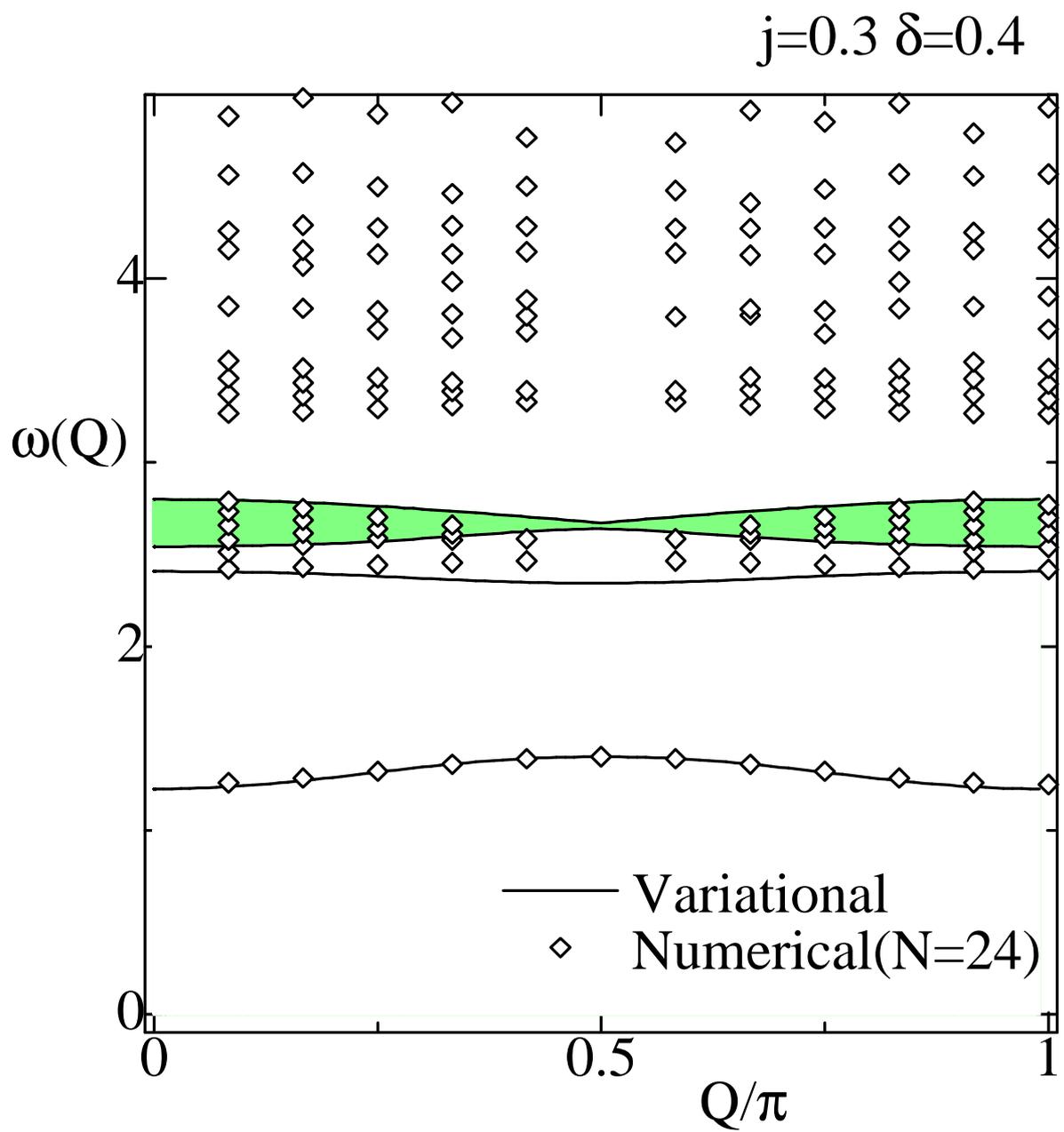
\item[Fig. 5]
Triplet excitation spectra for $j=0.3$ and $\delta=0.4$. The solid
 lines represent the result obtained by the variational calculation
 while the open diamonds denote the result obtained by the numerical
 diagonalization for the finite system ($N=24$). The shaded area denotes
 the continuum band obtained by the variational calculation.
\vspace*{1cm}
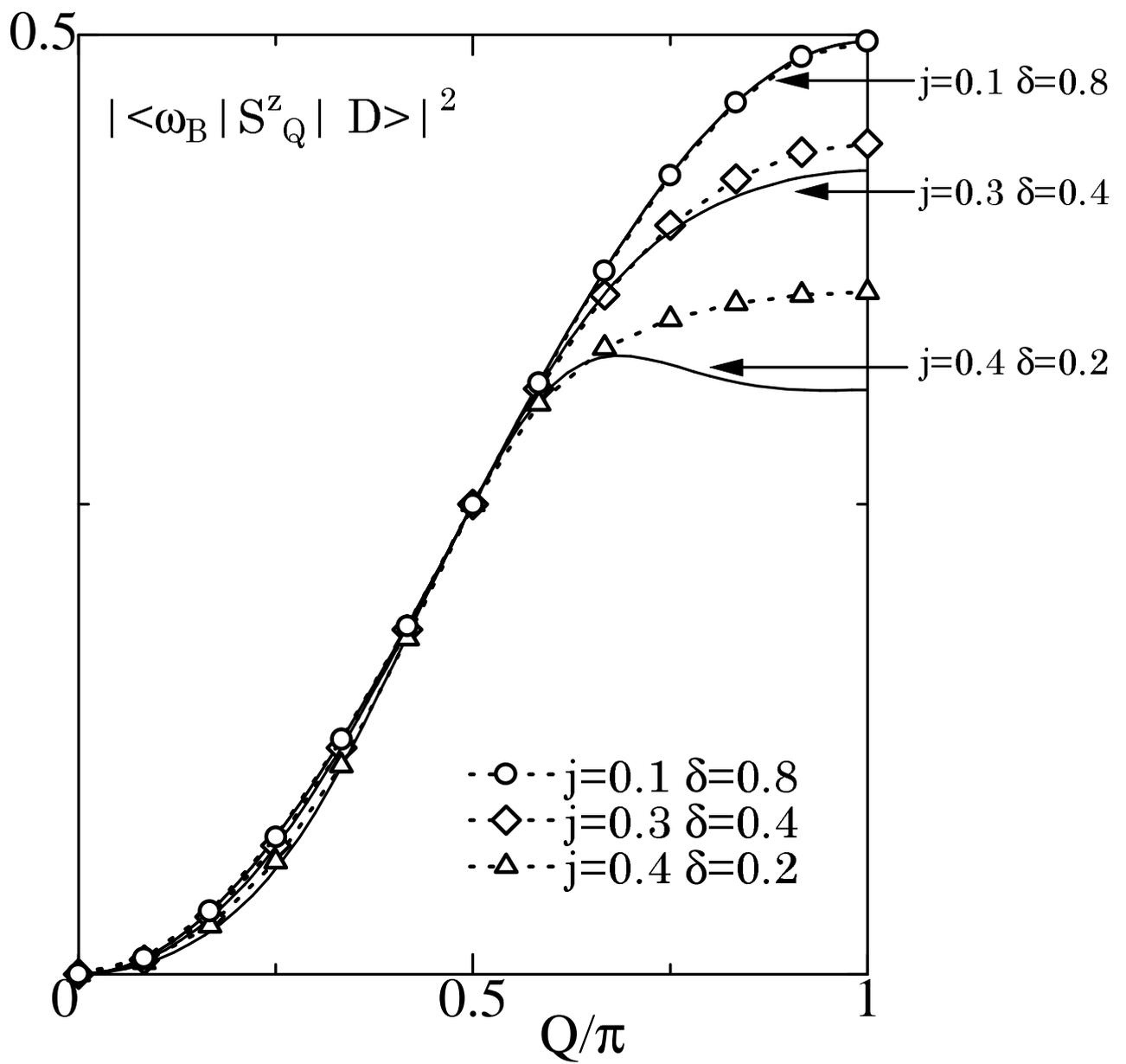
\item[Fig. 6]
Wavevector dependence of the spectral weight on the lowest excited mode
 $\omega_{\rm B}(Q)$. The numerical results of the finite chain ($N=24$)
 for $j=0.1$ and $\delta=0.8$, $j=0.3$ and $\delta=0.4$, and $j=0.4$ and
 $\delta=0.2$ are depicted with the circles, the diamonds, and the triangles,
 respectively. The broken lines are to guide the reader's eye. The solid lines
 denote the variational results.
\end{itemize}

\newpage

\section*{\large\bf References}

\begin{thebibliography}{99}
\bibitem{br} For review see, J. P. Boucher and L. P. Regnault: J. Phys. 
I (France) {\bf 6} (1996) 1939.
\bibitem{a} M. Arai , M. Fujita , M. Motokawa, J. Akimitsu 
and S. M. Bennington: Phys. Rev. Lett. {\bf 77} (1996) 3649.
\bibitem{al} M. A\"{\i}n, J. E. Lorenzo, L. P. Regnault,
 G. Dhalenne, A. Revcolevschi, B. Hennion and Th. Jolicoerur
: Phys. Rev. Lett. {\bf 78} (1997) 1560.
\bibitem{b} For review see, J. W. Bray, L. V. Interrante, I. S. Jacobs 
and J. C. Bonner: {\it Extended 
Linear Chain Compounds}, J. S. Miller ed. (Plenum Press, New York, 
1982) Vol. 3. p.353.
\bibitem{hmo} A preliminary result of the present work has been reported 
at ICM'97, Australia.
\bibitem{rd} J. Riera and A. Dobry: Phys. Rev. B {\bf 51} (1995) 16098.
\bibitem{c} G. Castilla , S. Chakravarty  and V. J. Emery
: Phys. Rev. Lett. {\bf 75} (1995) 1823.
\bibitem{cp} R. Chitra , S. Pati ,  H. R. Krishnamurthy, D. Sen 
and S. Ramasesha : Phys. Rev. {\bf B 52} (1995) 6581.
\bibitem{bm} S. Brehmer, H.-J. Mikeska and U. Neugebauer: J. Phys.; Cond. 
Mat. {\bf 8} (1996) 7161.
\bibitem{th} T. Tonegawa and I. Harada: J. Phys. Soc. Jpn. {\bf 56} 
(1987) 2153;  T. Tonegawa, I. Harada and M. Kaburagi
: {\it ibid.} {\bf 61} (1992) 4665.
\bibitem{on} K. Okamoto and K. Nomura
: Phys. Lett. A {\bf 169} (1992) 433.
\bibitem{ss} B. S. Shastry and B. Sutherland: Phys. Rev. Lett. {\bf 47} 
(1981) 964.
\bibitem{gb} E.R. Gagliano and C. A. Balseiro
: Phys. Rev. Lett. {\bf 59} (1987) 2999.
\bibitem{cm} W. J. Caspers and W. Magnus: Phys. Lett.
{\bf 88A} (1982) 103.
\bibitem{x}  We have calculated $\frac{1}{2}{\cal S}^{+-}(Q,\omega)$ 
rather than ${\cal S}^{zz}(Q,\omega)$ because of the convenience.  
It is, however, noted that these correlation functions are the same 
in the present isotropic system.
\bibitem{hk} I. Harada and A. Kotani: J. Phys. Soc. Jpn. {\bf 51} (1982) 
1737.
\bibitem{kk} V. Kiryukhin, B. Keimer, J. P. Hill and A. Vigliante:
 Phys. Rev. Lett. {\bf 76} (1996) 4608.
\bibitem{ys} H. Yokoyama and Y. Saiga: to be published in J. Phys. Soc. Jpn.
\end{thebibliography}
\end{document}